# Topological nature of higher-order hinge states revealed by spin transport

An-Qi Wang[1]†, Peng-Zhan Xiang[1]†, Tong-Yang Zhao[1], Zhi-Min Liao[1]*

[1]State Key Laboratory for Mesoscopic Physics and Frontiers Science Center for Nano-optoelectronics, School of Physics, Peking University; Beijing 100871, China.

*Corresponding author. Email: liaozm@pku.edu.cn

†These authors contributed equally to this work.

**Abstract:** One-dimensional (1D) gapless hinge states are predicated in the three-dimensional (3D) higher-order topological insulators and topological semimetals, because of the higher-order bulk-boundary correspondence. Nevertheless, the topologically protected property of the hinge states is still not demonstrated so far, because it is not accessible by conventional methods, such as spectroscopy experiments and quantum oscillations. Here, we reveal the topological nature of hinge states in the higher-order topological semimetal $Cd_3As_2$ nanoplate through spin potentiometric measurements. The results of current induced spin polarization indicate that the spin-momentum locking of the higher-order hinge state is similar to that of the quantum spin Hall state, showing the helical characteristics. The spin-polarized hinge states are robust up to room temperature and can nonlocally diffuse a long distance larger than 5 μm, further indicating their immunity protected by topology. Our work deepens the understanding of transport properties of the higher-order topological materials and should be valuable for future electronic and spintronic applications.

## 1. Introduction

Bulk-boundary correspondence that links the topological invariant of bulk band structure to boundary states is a unique characteristic of topological matters [1-5]. Typical bulk-boundary correspondence endows *d*-dimensional topological materials with gapless states at (*d-1*) dimensions (*i.e.*, the first-order topology). For example, a 3D "first-order" $Z_2$ topological insulator is featured with two-dimensional (2D) gapless helical surface states [1,2,6-11], while a 2D topological insulator with gapless boundary states at its 1D edges [12-15]. Recently, the bulk-boundary correspondence has been extended to higher-order cases enabling the lower-dimensional boundary states [16-31]. Particularly, a 3D second-order topological insulator possesses gapless modes at its 1D hinges, the intersection of two neighboring surface planes. Such second-order topological insulator phase has been indicated in bismuth [23], strained SnTe [22] and certain transition metal dichalcogenide materials [32-35]. Aside from the higher-order topological insulators, higher-order topological semimetals (for example, $Cd_3As_2$) are also proposed to host 1D hinge states [21,24-26,36]. Using the low-energy theory of quadrupole insulator, Wieder *et al.* captured the bulk-boundary correspondence of higher-order Dirac semimetal, predicting the 1D hinge Fermi arc as the direct topological consequence of bulk Dirac nodes [26]. Experimentally, the 1D conducting modes have been spatially resolved by scanning tunneling spectroscopy [23] and Josephson interferometry [23,33,37-40].

Despite extensive studies on its spatial identification, the topological nature of higher-order hinge state has still not been revealed so far. Moreover, the bulk and surface states are often in parallel conduction, rendering the 1D hinge modes elusive through conventional transport experiments. Here we reveal the topological nature and spin helicity of hinge states in the higher-order topological semimetal $Cd_3As_2$ nanoplate via spin potentiometric measurements.

## 2. Materials and methods

For higher-order Dirac semimetals, the quadruple-invariant protected hinge states span the projection of bulk 3D Dirac points along the 1D hinges [25,26,29], as sketched in Fig. 1a. For usually synthesized $Cd_3As_2$ nanoplates with (112) surface orientation, the 1D helical hinge states (Fig. 1b) are also expected due to the existence of projection along the (001) direction. The hinge states feature Kramers pairs of counterpropagating modes on opposite edges. In the presence of a current bias, the orientation of induced spin polarization is contrary on opposite edges of sample surface (Fig. 1c). When the current bias is reversed, the induced spin polarization of each edge would also be reversed.

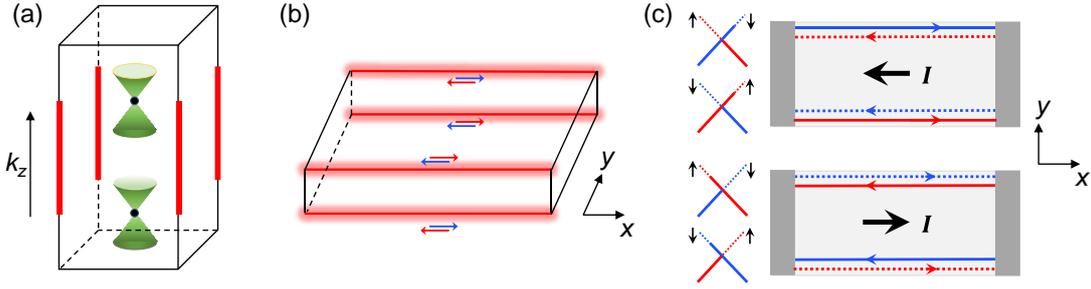

**Fig. 1. Helical hinge states in the higher-order topological semimetal.**

(a) Schematic of hinge states of higher-order Dirac semimetal in momentum space. Each hinge mode spans the projection of two bulk Dirac points on the hinge.

(b) Hinge states distributed on the top and bottom surface of a $Cd_3As_2$ film with (112) surface orientation. $Cd_3As_2$ has two Dirac points along $k_z$ direction, *i.e.*, (001) direction.

(c) Current-induced spin polarizations for hinge states. The diagram to the left depicts the population of hinge states, where the blue (red) branch corresponds to spin-down (up) electrons. The current induced spin-polarization directions are opposite for the two neighboring hinges. As changing the polarity of bias current, the spin-polarization directions are reversed accordingly for each hinge.

The employed $Cd_3As_2$ nanoplates with (112) surface orientation (Supplementary materials, Fig. S1) were grown by chemical vapor deposition [41,42]. Individual nanoplates were then transferred onto a silicon substrate with a 285-nm-thick $SiO_2$ layer, which serves as back gate to tune the Fermi level of $Cd_3As_2$. Co electrodes were used for spin detection with the Ti/Au electrodes as reference. To compare the spin transport properties on the opposite sides, pairs of Hall-bar-like Co electrodes were fabricated on the two edges of $Cd_3As_2$ nanoplate. A 3-nm-thick $Al_2O_3$ oxide layer was inserted between Co and $Cd_3As_2$. On one hand, the existence of tunnel barrier can overcome the conductance mismatch problem [43] and thus enhance the spin detection efficiency [44-46], which has been widely applied to detect current-induced spin polarization of topological materials [6,47-51]. On the other hand, the $Al_2O_3$ oxide layer can protect the hinge states from magnetic scatterings of Co electrodes. A total of twenty-two devices have been investigated in our work, all showing consistent spin transport properties (Supplementary materials, Fig. S2). Here, we mainly presented the results from four typical devices, marked as device A, B, C and D.

## 3. Result and discussion

### 3.1. Spin-momentum locking of the hinge states

Figure 2 shows the spin transport results measured on the device A. As sketched in Fig. 2a, a direct current (dc) was injected into the nanoplate via the outmost two Ti/Au electrodes, and the voltage was measured between the inner Co and Ti/Au electrodes. An external magnetic field was applied along the long axis of the cobalt strip, in-plane perpendicular to the current direction, to modulate the Co magnetization $M$ for probing the spin signals. The As shown in Fig. 2b, c, with a bias current 80 μA, magnetic hysteretic loops are clearly observed on both two sides of the nanoplate sample. The spin-dependent voltage $V_{1S}$ and $V_{2S}$ are obtained after subtracting the background from raw data (Supplementary materials, Fig. S3), where the subscript 1 and 2 are employed to distinguish the upper and lower edges.

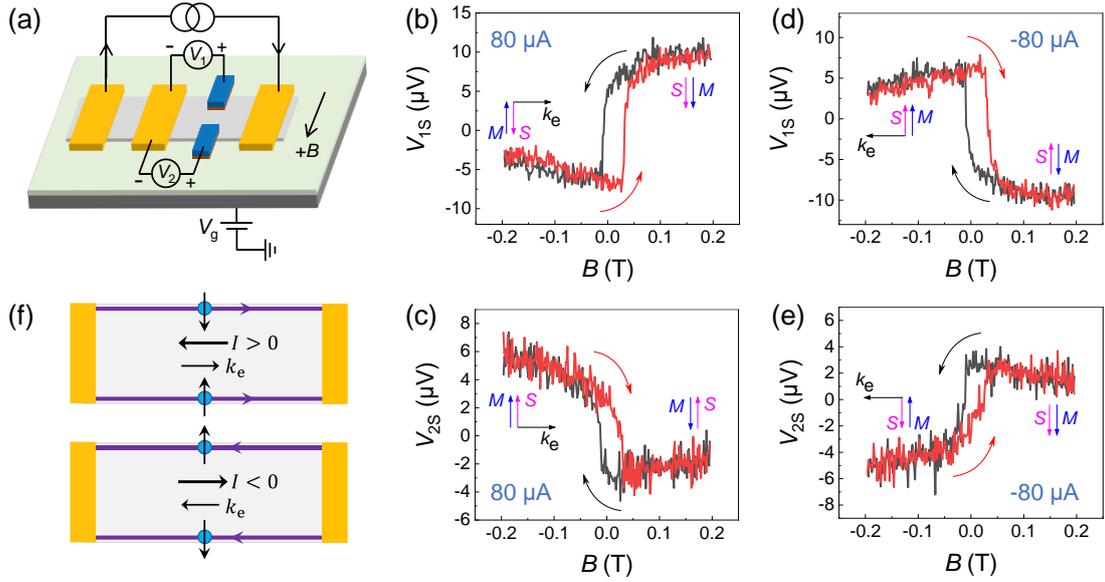

**Fig. 2. Spin transport on the two edges of nanoplate device A.**

(a) Schematic of the device structure and measurement configuration. Ti/Au and Co electrodes are denoted by yellow and blue, respectively. A 3-nm-thick $Al_2O_3$ layer is inserted between the Co and $Cd_3As_2$. The gate voltage $V_g$ is applied onto the Si substrate for tuning the sample Fermi level.

(b)-(e) Spin transport measured on the upper edge (b, d) and the lower edge (c, e) with bias current $I = \pm 80$ µA at 1.4 K. Insets show the corresponding spin-momentum locking. The red and black arrows denote the forward and backward sweeping of magnetic fields, respectively.

(f) Schematic of the current-induced opposite spin polarizations for the two edges. The spin $S$ is locked to electron momentum $k_e$ with right-hand and left-hand rules for the upper and lower edges, respectively. The blue filled circles represent the electrons and the black arrows above denote the spin direction. The purple arrows show the motion of electrons on the edge.

As sweeping the magnetic field $B$, a counterclockwise hysteretic loop is captured on the upper edge of nanoplate, while a clockwise hysteretic loop on the lower edge. The direction of spin polarization $S$ can be determined according to the measured voltage $V_S$ and the magnetization $M$ of the Co electrode. Specifically, $S//M$ and $S//-M$ result in the high and low voltage states, respectively [46,52]. A dc bias generally gives rise to a net momentum $k_\text{e}$ in the electronic system. With the determined orientation $S$ and $k_\text{e}$, we can obtain the right-handed and left-handed spin-momentum locking on the upper and lower edges, respectively, as seen in the insets of Fig. 2b, c. For each edge, upon reversing the current direction, the spin polarization also reverses its direction (Fig. 2d, e), further confirming the spin-momentum locking feature. Consistent results are also obtained in other nanoplate devices (see the results of different devices in Supplementary materials, Figs. S4-S6). The measured opposite spin directions (Fig. 2f) on the two sides of a single nanoplate cannot stem entirely from the surface state transport [6,41,42,46-50], because the current-induced spin polarization of the surface states should be spatially independent [49,53]. The detailed comparisons of surface state and hinge state spin transport are described in Supplementary materials Fig. S7. The observed spin transport properties on the two edges are consistent with existence of higher-order hinge states in $Cd_3As_2$. Other mechanisms, such as, Rashba states [46], side wall surface states, and quantum spin Hall effect, all fail to explain our experimental results (see the discussion part in Supplementary materials).

We further perform spin detection measurements on a single device via two types of spin probes, namely, the Co electrodes only contacting the one side of nanoplate (Hall-bar-like) and the ones across the whole nanoplate (Fig. 3a). Similar to device A, opposite spin polarization directions are detected on one pair of Hall-bar-like Co electrodes for the two sides of device B, as shown in Fig. 3b, c. Driven by a current bias $I = 40$ μA, the spin voltage signal $V_{1S}$ and $V_{2S}$ have similar loop heights of about 10 μV. We define the hysteretic window height as $\Delta V_S$, and the spin resistance as $\Delta R_S = \Delta V_S/I$, where $\Delta V_S$ is positive for clockwise magnetic loops and negative for counterclockwise loops. The $|\Delta R_S|$ is about 0.25 Ω for a single hinge in device B, which is greatly underestimated due to the edge current accounting for only a small part of the total current. The $\Delta V_S$ is also found linearly proportional to $I$ (Fig. 3d), indicating that the spin polarization ratio is independent of bias current [47,52]. Spin related potential difference between the two opposite Co electrodes is also measured (Fig. 3e). The obtained $|\Delta V_{1-2,S}| \approx 22$ μV, approximately the value of $|\Delta V_{1S}| + |\Delta V_{2S}|$, further confirming the opposite spin polarization directions of the two edges. When the Co electrode is across the nanoplate (Fig. 3f), the spin polarizations from opposite hinges would be greatly cancelled out, recalling the surface spin transport. The cancellation renders the decrease of charge-to-spin conversion efficiency, as presented in Fig. 3f, where the $|\Delta R_{3S}|\sim 0.03$ Ω is only one eighth of that from one edge. The inset of Fig. 3f demonstrates the spin texture of Rashba surface states, in agreement with previous studies as the charge transfer from the Co electrode can enhance the interfacial electric field [41]. Besides the surface state contribution, the inequivalence of hinge states at two edges would also provide a possible origin for the nonzero spin signals in Fig. 3(f).

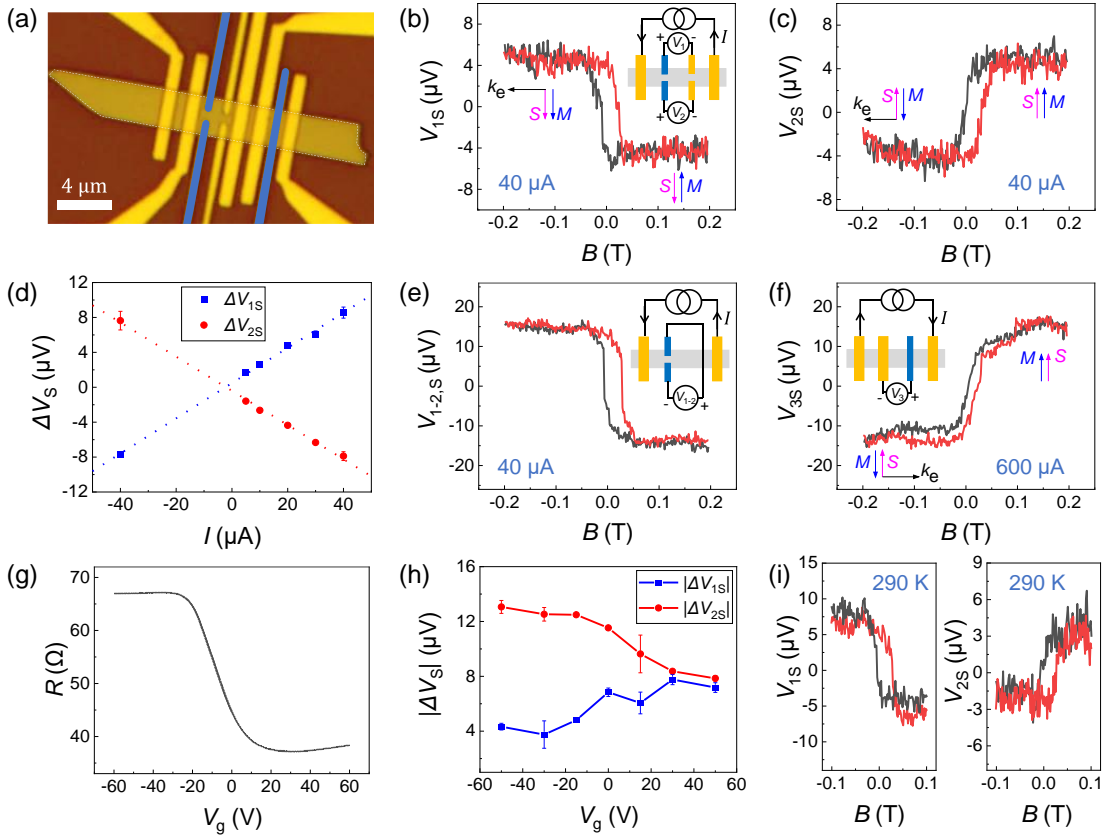

**Fig. 3. Influence of electrode configuration, gate voltage and temperature on the edge spin transport.**

(a) Optical image of the device B. The Co and Ti/Au electrodes are denoted by blue and yellow, respectively.

(b)-(c) Spin voltage $V_{1S}$ and $V_{2S}$ loop measured on the two sides of device B, respectively at 1.4 K. The applied bias current is $I = 40\ \mu A$. The measurement configuration is shown by the inset of (b). Left-handed spin-momentum locking is detected on the upper edge, while right-handed spin-momentum locking on the lower edge, indicated by the insets.

(d) The magnetic hysteresis window height $\Delta V_S$ as a function of bias current $I$ for the upper and lower edges of device B at 1.4 K. Dashed lines are linear fits to the experimental data. The error bars represent the standard deviation over multiple measurements.

(e) Spin voltage $V_{1-2,S}$ as a function of magnetic fields in device B at 1.4 K. $V_{1-2}$ is the voltage drop between the two Hall-bar-like Co electrodes, while the $V_{1-2,S}$ is obtained after subtracting the background. The applied bias current is $I = 40\ \mu A$ and the measurement configuration is illustrated by the inset.

(f) Spin voltage $V_{3S}$ measured as the Co electrode is across the sample in device B at 1.4 K. The applied bias current is $I = 600\ \mu A$. The inset depicts the measurement configuration.

(g) The transfer curve of device C, obtained from the standard four-probe measurements at 1.4 K. Device C has nearly the same contact geometry as that of device B.

(h) Gate voltage dependence of magnetic hysteresis window height $|\Delta V_{1S}|$ and $|\Delta V_{2S}|$ for two opposite edges in device C at 1.4 K. The applied current bias $I = 40\ \mu A$. The error bars represent the standard deviation over multiple measurements.

(i) Spin transport at 290 K, measured on the upper edge (left panel) and lower edge (right panel) of device C.

The relative contribution of surface and hinge spin signals can be tuned not only by different electrode configurations, but also by the gate voltage. The transfer curve in Fig. 3g shows the Dirac point ~ -30 V in the nanoplate of device C, and the gate voltage dependence of spin-polarized transport on the two opposite sides is shown in Fig. 3h. The $|\Delta V_S|$ from the two sides demonstrates opposite tendency with varying gate voltage $V_g$. For the contributions from surface states, it decreases as tuning $V_g$ away from the Dirac point due to the decrease of surface-bulk conduction ratio [41,42,52]. The opposite spin signals of the two hinges are combined with the contribution of surface states, so that the total measurement signal is strengthened on one side and weakened on the other. As the $V_g$ is tuned to a large positive value, the surface spin contribution can be ignored, and nearly identical $|\Delta V_S|$ are observed on two sides for $V_g = 50$ V (Fig. 3h). Furthermore, the edge spin signals are robust against temperature. As presented in Fig. 3i, the spin signals on the two edges can both exist steadily at 290 K, holding promise for the practical spintronic devices in the future.

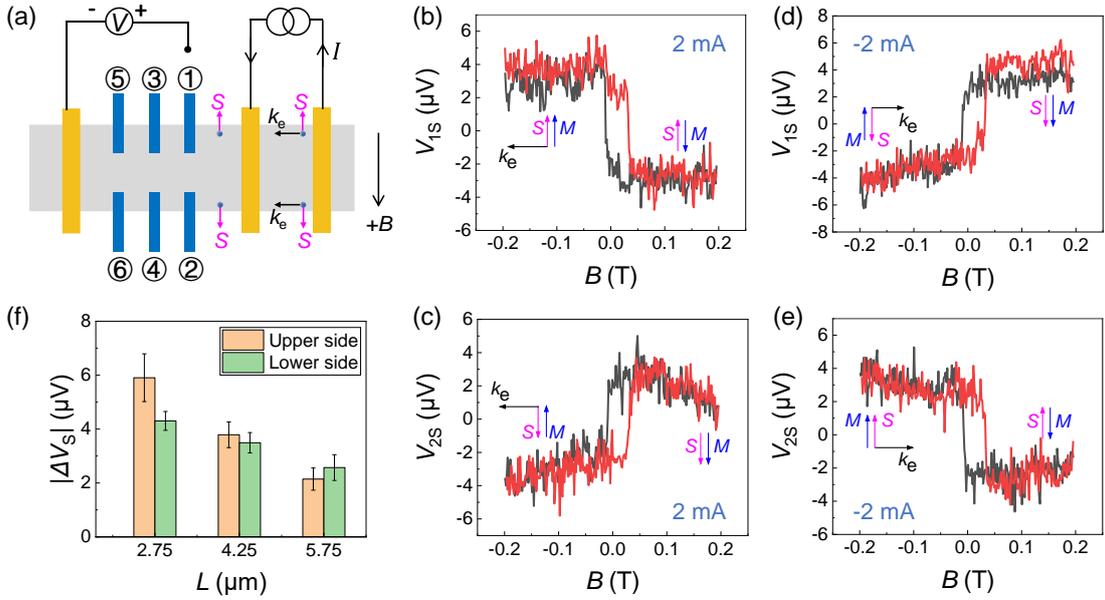

**Fig. 4. Nonlocal spin transport on opposite sides of device A at 1.4 K.**

(a) Diagram of nonlocal measurement configuration. The nonlocal voltage measured on Co electrode 1-6 is denoted by $V_1$, $V_2$, …, $V_6$, respectively. The bias current locally induces opposite spin polarizations on the upper and lower edges, and the spin diffusion results in the nonlocal signals.

(b)-(e) Nonlocal spin transport measured on the upper edge (b, d) and the lower edge (c, e) with bias current $I = \pm 2$ mA. Insets show the corresponding spin-momentum locking, where the spin direction $S$ is determined according to the measured voltage $V_S$ and the magnetization *M* of the Co electrode.

(f) The $|\Delta V_S|$ as a function of distance *L*, where *L* is the spacing length between the nonlocal Co electrode and its nearest local Ti/Au electrode. The Co electrodes 1 (2), 3 (4) and 5 (6) correspond to a distance of 2.75, 4.25 and 5.75 μm, respectively. The error bars represent standard deviation over multiple measurements.

## *3.2. Nonlocal spin transport on the hinges*

Driven by the dc bias, the electrons of edge states acquire net spin polarization in the local region. The spin diffusion process allows us to detect the spin-related hysteretic loops in the region away from the local source electrode. As sketched in Fig. 4a, the dc bias is applied to the nanoplate via the rightmost two Ti/Au electrodes, while the voltage is measured nonlocally between the Co electrode and the leftmost Ti/Au electrode. We here fabricate several pairs of Hall-bar-like Co electrodes, denoted by the Roman number 1 to 6, serving as nonlocal spin detectors. When applying a bias current $I = 2$ mA, hysteretic loops are clearly observed on the upper and lower edges of the nanoplate (Fig. 4b, c). Similar to the local measurement results, these two edges still demonstrate opposite spin-polarized signals in the nonlocal regime. Upon changing the polarity of the bias current, the spin polarization of the nonlocal transport also experiences a reversal (Fig. 4d, e), showing consistent spin-momentum locking with the local channel. To suppress the charge-current-spreading effect, the nearest nonlocal spin probe is set beyond 2 μm from the local electrode, where the distance is larger than the mean free path of electrons [54-56], facilitating the detection of spin diffusion process (Supplementary materials, Fig. S6). Figure 4f shows the $|\Delta V_S|$ as a function of *L*, where *L* is the distance between the nonlocal spin detector Co and the its nearest local Ti/Au electrode. Although the nonlocal spin signals generally decay with increasing *L*, the spin-polarized signals can still be unambiguously detected at a considerable distance of $L = 5.75$ μm. The long-distance nonlocal spin diffusion should result from the topological protection of hinge states, which can greatly suppress the backscattering and spin flipping.

## 4. Conclusions

In summary, we have identified the helical hinge states in the higher-order topological semimetal by employing spin potentiometric measurements. Benefiting from topological protection and the strict prohibition of backscattering of non-magnetic impurities in the 1D hinge channel, the current induced spin polarization is still observable at room temperature and the nonlocal spin diffusion length is longer than 5 μm. Our work should be valuable for understanding the transport properties of higher-order topological states. The topological edge transport channels are easy to replicate, integrate and scale up, paving the way for low-dissipation electronic and spintronic devices.

## Conflict of interest

The authors declare that they have no conflict of interest.

## Acknowledgements

This work was supported by National Natural Science Foundation of China (Grant Nos. 91964201 and 61825401) and China Postdoctoral Science Foundation (Grant No. 2021M700254).

## Author contributions

Zhi-Min Liao conceived and supervised this work. Peng-Zhan Xiang, An-Qi Wang and Tong-Yang Zhao fabricated the devices and performed the measurements. Zhi-Min Liao, An-Qi Wang and Peng-Zhan Xiang analyzed the data. An-Qi Wang and Zhi-Min Liao wrote the paper.

## Appendix A. Supplementary materials

Supplementary materials to this article can be found online at ×××××××××.

# References

[1] Hasan MZ, Kane CL. Colloquium: Topological insulators. Rev Mod Phys 2010;82:3045-3067.

[2] Qi X-L, Zhang S-C. Topological insulators and superconductors. Rev Mod Phys 2011;83:1057-1110.

[3] Chiu C-K, Teo JCY, Schnyder AP, et al. Classification of topological quantum matter with symmetries. Rev Mod Phys 2016;88:035005.

[4] Moore JE, Balents L. Topological invariants of time-reversal-invariant band structures. Phys Rev B 2007;75:121306.

[5] Armitage NP, Mele EJ, Vishwanath A. Weyl and Dirac semimetals in three-dimensional solids. Rev Mod Phys 2018;90:015001.

[6] Li CH, van 't Erve OMJ, Robinson JT, et al. Electrical detection of charge-current-induced spin polarization due to spin-momentum locking in $Bi_2Se_3$. Nat Nanotechnol 2014;9:218-224.

[7] Mellnik AR, Lee JS, Richardella A, et al. Spin-transfer torque generated by a topological insulator. Nature 2014;511:449-451.

[8] Wang Y, Deorani P, Banerjee K, et al. Topological surface states originated spin-orbit torques in $Bi_2Se_3$. Phys Rev Lett 2015;114:257202.

[9] Han J, Richardella A, Siddiqui SA, et al. Room-temperature spin-orbit torque switching induced by a topological insulator. Phys Rev Lett 2017;119:077702.

[10] Fan Y, Upadhyaya P, Kou X, et al. Magnetization switching through giant spin–orbit torque in a magnetically doped topological insulator heterostructure. Nat Mater 2014;13:699-704.

[11] Han J, Liu L. Topological insulators for efficient spin–orbit torques. APL Mater 2021;9:060901.

[12] Kane CL, Mele EJ. $Z_2$ Topological order and the quantum spin Hall effect. Phys Rev Lett 2005;95:146802.

[13] Bernevig BA, Zhang S-C. Quantum spin Hall effect. Phys Rev Lett 2006;96:106802.

[14] Bernevig BA, Hughes TL, Zhang S-C. Quantum spin Hall effect and topological phase transition in HgTe quantum wells. Science 2006;314:1757-1761.

[15] König M, Wiedmann S, Brüne C, et al. Quantum spin Hall insulator state in HgTe quantum wells. Science 2007;318:766-770.

[16] Benalcazar WA, Bernevig BA, Hughes TL. Quantized electric multipole insulators. Science 2017;357:61-66.

[17] Benalcazar WA, Bernevig BA, Hughes TL. Electric multipole moments, topological multipole moment pumping, and chiral hinge states in crystalline insulators. Phys Rev B 2017;96:245115.

[18] Song Z, Fang Z, Fang C. (D−2)-dimensional edge states of rotation symmetry protected topological states. Phys Rev Lett 2017;119:246402.

[19] Imhof S, Berger C, Bayer F, et al. Topolectrical-circuit realization of topological corner modes. Nat Phys 2018;14:925-929.

[20] Khalaf E. Higher-order topological insulators and superconductors protected by

inversion symmetry. Phys Rev B 2018;97:205136.
[21] Lin M, Hughes TL. Topological quadrupolar semimetals. Phys Rev B 2018;98:241103.
[22] Schindler F, Cook AM, Vergniory MG, et al. Higher-order topological insulators. Sci Adv 2018;4:eaat0346.
[23] Schindler F, Wang Z, Vergniory MG, et al. Higher-order topology in bismuth. Nat Phys 2018;14:918-924.
[24] Ezawa M. Higher-Order Topological insulators and semimetals on the breathing Kagome and Pyrochlore lattices. Phys Rev Lett 2018;120:026801.
[25] Călugăru D, Juričić V, Roy B. Higher-order topological phases: A general principle of construction. Phys Rev B 2019;99:041301.
[26] Wieder BJ, Wang Z, Cano J, et al. Strong and fragile topological Dirac semimetals with higher-order Fermi arcs. Nat Commun 2020;11:627.
[27] Ezawa M. Magnetic second-order topological insulators and semimetals. Phys Rev B 2018;97:155305.
[28] Ezawa M. Second-order topological insulators and loop-nodal semimetals in transition metal dichalcogenides $XTe_2$ (X = Mo, W). Sci Rep 2019;9:5286.
[29] Wei Q, Zhang X, Deng W, et al. Higher-order topological semimetal in acoustic crystals. Nat Mater 2021;20:812-817.
[30] Langbehn J, Peng Y, Trifunovic L, et al. Reflection-symmetric second-order topological insulators and superconductors. Phys Rev Lett 2017;119:246401.
[31] Trifunovic L, Brouwer PW. Higher-order bulk-boundary correspondence for topological crystalline phases. Phys Rev X 2019;9:011012.
[32] Wang Z, Wieder BJ, Li J, et al. Higher-order topology, monopole nodal lines, and the origin of large Fermi arcs in transition metal dichalcogenides $XTe_2$ (X = Mo, W). Phys Rev Lett 2019;123:186401.
[33] Choi Y-B, Xie Y, Chen C-Z, et al. Evidence of higher-order topology in multilayer $WTe_2$ from Josephson coupling through anisotropic hinge states. Nat Mater 2020;19:974-979.
[34] Wang W, Kim S, Liu M, et al. Evidence for an edge supercurrent in the Weyl superconductor $MoTe_2$. Science 2020;368:534-537.
[35] Wang Y, Lee G-H, Ali MN. Topology and superconductivity on the edge. Nat Phys 2021;17:542-546.
[36] Chen R, Liu T, Wang CM, et al. Field-tunable one-sided higher-order topological hinge states in Dirac semimetals. Phys Rev Lett 2021;127:066801.
[37] Murani A, Kasumov A, Sengupta S, et al. Ballistic edge states in Bismuth nanowires revealed by SQUID interferometry. Nat Commun 2017;8:15941.
[38] Kononov A, Abulizi G, Qu K, et al. One-dimensional edge transport in few-layer $WTe_2$. Nano Lett 2020;20:4228-4233.
[39] Li C-Z, Wang A-Q, Li C, et al. Reducing electronic transport dimension to topological hinge states by increasing geometry size of Dirac semimetal Josephson junctions. Phys Rev Lett 2020;124:156601.
[40] Huang C, Narayan A, Zhang E, et al. Edge superconductivity in multilayer $WTe_2$ Josephson junction. Natl Sci Rev 2020;7:1468-1475.

[41] Wang A-Q, Xiang P-Z, Ye X-G, et al. Room-temperature manipulation of spin texture in a Dirac semimetal. Phys Rev Applied 2020;14:054044.

[42] Wang A-Q, Xiang P-Z, Ye X-G, et al. Surface engineering of antisymmetric linear magnetoresistance and spin-polarized surface state transport in Dirac semimetals. Nano Lett 2021;21:2026-2032.

[43] Schmidt G, Ferrand D, Molenkamp LW, et al. Fundamental obstacle for electrical spin injection from a ferromagnetic metal into a diffusive semiconductor. Phys Rev B 2000;62:R4790-R4793.

[44] Rashba EI. Theory of electrical spin injection: Tunnel contacts as a solution of the conductivity mismatch problem. Phys Rev B 2000;62:R16267-R16270.

[45] Fert A, Jaffrès H. Conditions for efficient spin injection from a ferromagnetic metal into a semiconductor. Phys Rev B 2001;64:184420.

[46] Yang F, Ghatak S, Taskin AA, et al. Switching of charge-current-induced spin polarization in the topological insulator $BiSbTeSe_2$. Phys Rev B 2016;94:075304.

[47] Dankert A, Geurs J, Kamalakar MV, et al. Room temperature electrical detection of spin polarized currents in topological insulators. Nano Lett 2015;15:7976-7981.

[48] Tang J, Chang L-T, Kou X, et al. Electrical detection of spin-polarized surface states conduction in $(Bi_{0.53}Sb_{0.47})_2Te_3$ topological insulator. Nano Lett 2014;14:5423-5429.

[49] Tian J, Miotkowski I, Hong S, et al. Electrical injection and detection of spin-polarized currents in topological insulator $Bi_2Te_2Se$. Sci Rep 2015;5:14293.

[50] Lin B-C, Wang S, Wang A-Q, et al. Electric control of Fermi arc spin transport in individual topological semimetal nanowires. Phys Rev Lett 2020;124:116802.

[51] Li P, Wu W, Wen Y, et al. Spin-momentum locking and spin-orbit torques in magnetic nano-heterojunctions composed of Weyl semimetal $WTe_2$. Nat Commun 2018;9:3990.

[52] Hong S, Diep V, Datta S, et al. Modeling potentiometric measurements in topological insulators including parallel channels. Phys Rev B 2012;86:085131.

[53] Tian J, Şahin C, Miotkowski I, et al. Opposite current-induced spin polarizations in bulk-metallic $Bi_2Se_3$ and bulk-insulating $Bi_2Te_2Se$ topological insulator thin flakes. Phys Rev B 2021;103:035412.

[54] Hwang T-H, Kim H-S, Kim H, et al. Electrical detection of spin-polarized current in topological insulator $Bi_{1.5}Sb_{0.5}Te_{1.7}Se_{1.3}$. Curr Appl Phys 2019;19:917-923.

[55] Hu J, Stecklein G, Deen DA, et al. Scaling of the nonlocal spin and baseline resistances in graphene lateral spin valves. IEEE Trans Electron 2019;66:5003-5010.

[56] Johnson M, Silsbee RH. Calculation of nonlocal baseline resistance in a quasi-one-dimensional wire. Phys Rev B 2007;76:153107.

Supplementary materials for:

# Topological nature of higher-order hinge states revealed by spin transport

An-Qi Wang[1]†, Peng-Zhan Xiang[1]†, Tong-Yang Zhao[1], Zhi-Min Liao[1]*

[1]State Key Laboratory for Mesoscopic Physics and Frontiers Science Center for Nano-optoelectronics, School of Physics, Peking University; Beijing 100871, China.

*Corresponding author. Email: liaozm@pku.edu.cn

†These authors contributed equally to this work.

**Sample growth and transport measurements**

Cd$_3$As$_2$ nanoplates were grown by chemical vapor deposition with (112) surface orientation. Cd$_3$As$_2$ powders with high purity (> 99.99%) were placed in the center of horizontal quartz tube. Silicon wafers with 5 nm gold thin film were placed downstream as substrates to collect the products. The quartz tube was first flushed three times with Argon gas to get rid of oxygen, then gradually heated from room temperature to 700 $^0$C within 15 minutes, and kept for 10 minutes at 700 $^0$C along with an Argon gas flow of 20 s.c.c.m. The system was then cooled down naturally. Cd$_3$As$_2$ nanoplates were collected on the silicon wafer substrates. **Figure S1a** shows the scanning electron microscopy (SEM) image of the as-grown nanoplates. The selected area electron diffraction (SAED) pattern is shown in **Fig. S1b**. According to the crystallography calculation and hexagonal lattice symmetry, (112) surface orientation can be determined for the nanoplate.

Individual Cd$_3$As$_2$ nanoplates were then transferred onto a silicon substrate with a 285-nm-thick oxide layer. The thickness of selected nanoplate is about 100 nm. Ti/Au and Co/Au electrodes were fabricated after two rounds of e-beam lithography and e-beam evaporation process. A gold layer above was used as a capping layer to protect Co from oxidization. To establish the Ohmic contact between Cd$_3$As$_2$ and Ti/Au electrode, an in-situ Ar$^+$ etching process was performed to remove the native oxide layer of the nanoplate before metal deposition. A 3-nm-thick Al$_2$O$_3$ layer was inserted between Cd$_3$As$_2$ and Co electrode using e-beam evaporation system. **Figure S1c** presents the optical image of a finished nanoplate device.

Transport measurements were performed in a commercial Oxford cryostat system with a base temperature ~ 1.4 K. The magnetic field was aligned along the easy axis of Co strip, in-plane perpendicular to the current direction. The devices were measured with a Keithley 2400 Source Meter using direct current. For the four-probe spin potentiometric measurement, the current was injected into the nanoplate via source and drain (Ti/Au electrodes), measuring the voltage drop between the Co and another Ti/Au electrode with the sweeps of magnetic fields.

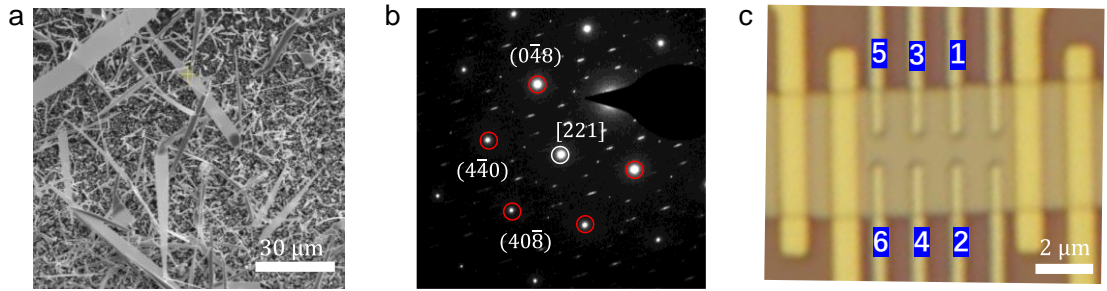

**Figure S1: Characterization of the synthesized Cd$_3$As$_2$ nanoplates and typical nanoplate device.**

**a**, SEM image of the as-grown nanoplates.

**b**, The selected electron diffraction (SAED) pattern of the nanoplate. By calculating the distance from the diffraction point to the center point and their relative orientations, $(0\bar{4}8)$, $(4\bar{4}0)$ and $(40\bar{8})$ crystal planes are determined as denoted. Based on the crystallography calculation, it indicates that the nanoplate has the (112) top surface plane with [221] zone axis.

**c**, Optical image of device A in the main text. The wider electrodes correspond to the Ti/Au electrodes, while the narrower ones denote the Co electrodes. The Co electrodes marked by 1-6 are consistent with the numbering in Fig. 4 of the main text.

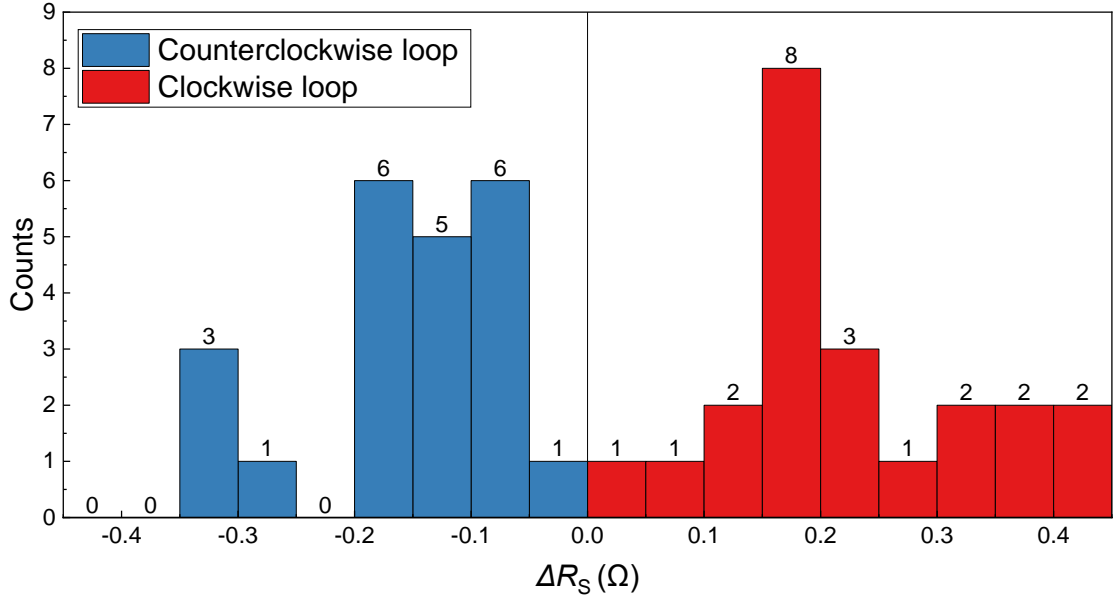

**Figure S2: Statistics of the measured devices showing opposite spin polarizations on opposite edges at 1.4 K.** Spin resistance is defined as $\Delta R_S = \Delta V_S/I$, where $\Delta V_S$ is the magnetic loop height as sweeping the magnetic field to change the magnetization direction of the Co electrode. Here, the negative value of $\Delta R_S$ corresponds to the edge with counterclockwise magnetic loop, while positive one denotes the other edge with clockwise magnetic loop.

A total of 22 devices have been investigated in our work. In all of these devices, opposite spin-polarized signals are observed on the two sides of nanoplate. **Figure S2** gives the counting results of these devices in terms of the spin resistance $\Delta R_S = \Delta V_S/I$, which can be employed to characterize the charge-to-spin conversion efficiency. The positive and negative $\Delta R_S$ correspond to the clockwise and counterclockwise magnetic loops, respectively. Limited by the process of $Ar^+$ milling and $Al_2O_3$ deposition, the $Co/Cd_3As_2$ interface would greatly affect the spin detection efficiency. Moreover, carrier density varies for different nanoplates. Both effects together lead to the device-to-device variations in the spin resistance.

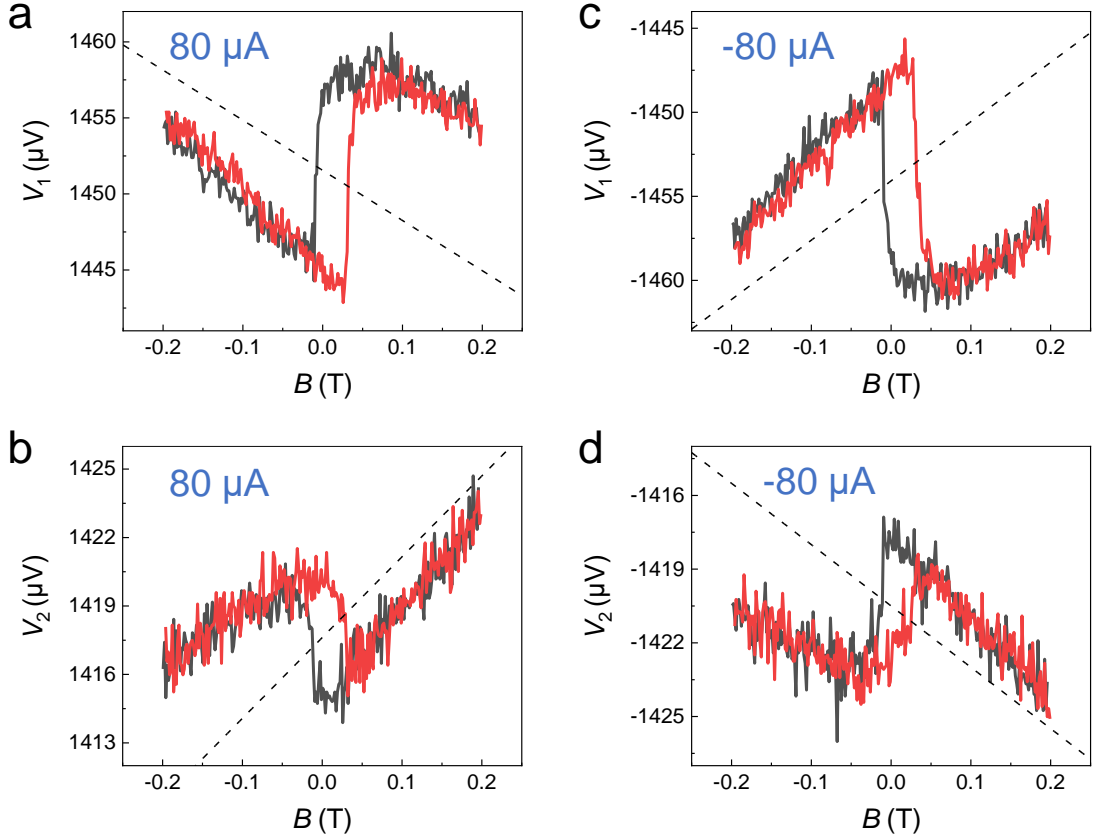

**Figure S3：Raw data of spin transport measured on the two edges in device A at 1.4 K.**

**a, c,** Raw data of magnetic hysteretic loops measured on the upper edge, corresponding to Fig. 2b and Fig. 2d in the main text, respectively. The applied bias current is $I = \pm 80$ μA. The background is denoted by dashed lines.

**b, d,** Raw data of magnetic hysteretic loops measured on the lower edge, corresponding to Fig. 2c and Fig. 2e in the main text, respectively. The applied bias current is $I = \pm 80$ μA. The background is denoted by dashed lines.

**Figure S3** shows the raw data of spin transport measured on device A, where the linear magnetoresistance background is clearly observed (marked by dashed lines). The linear background has been previously investigated [1,2], which may originate from the vertical charge transport between the bulk and surface conduction channels. After subtracting the linear background, we can obtain the spin-dependent voltage $V_S$ versus magnetic field $B$.

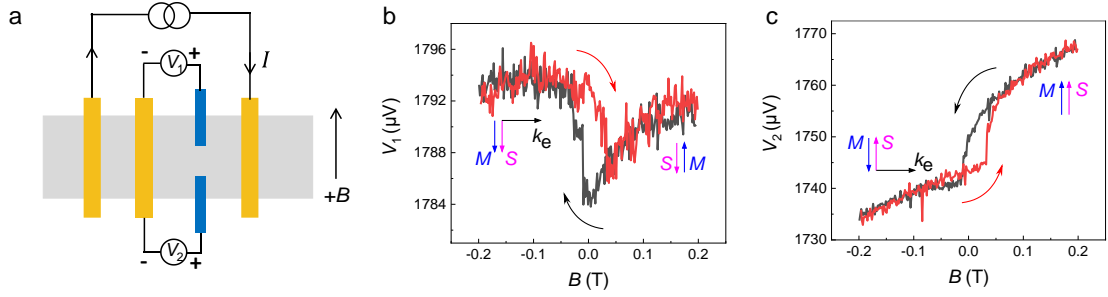

**Figure S4: Raw data of local spin transport measured on the two edges in device C at 1.4 K.**

**a**, Schematic of local spin detection measurements. The voltage measured on the upper and lower edge is denoted by $V_1$ and $V_2$, respectively.

**b,** Raw data of magnetic hysteretic loops measured on the upper edge. The applied bias current is $I = 40$ µA. The spin $S$ is locked to electron momentum $k_e$ with right-hand rule for the upper edge, as depicted by the inset.

**c,** Raw data of magnetic hysteretic loops measured on the lower edge. The applied bias current is $I = 40$ µA. The spin $S$ is locked to electron momentum $k_e$ with left-hand rule for the lower edge, as depicted by the inset.

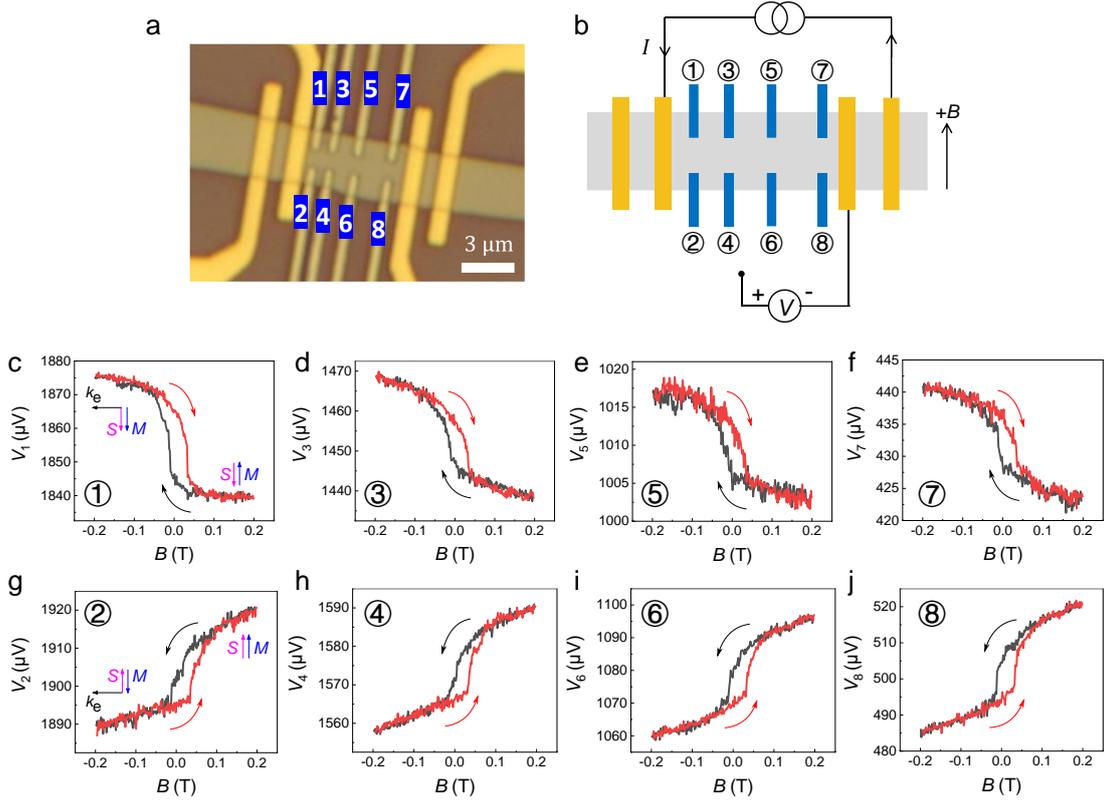

**Figure S5: Raw data of local spin transport measured on the two edges in device D at 1.4 K.**

**a**, Optical image of nanoplate device D. Four pairs of Hall-bar-like Co electrodes are fabricated and denoted by numbers 1 to 8.

**b,** Schematic diagram of local spin detection measurements. Upon applying a bias current, the voltage is measured on each Co electrode with Ti/Au electrode as reference. The voltage measured on Co electrode 1-8 is denoted by $V_1$, $V_2$, …, $V_8$, respectively.

**c-f,** Magnetic hysteretic loops measured on different positions of the upper edge. The applied bias current is $I = 60$ μA. Left-handed spin-momentum locking is detected on this edge.

**g-j,** Magnetic hysteretic loops measured on different positions of the lower edge. The applied bias current is $I = 60$ μA. Right-handed spin-momentum locking is detected on this edge.

For a given edge of nanoplate, we find the orientation of current-induced spin polarization is spatially independent and remains unchanged upon different positions.

Similar to device A discussed in the main text, nanoplate device D also demonstrates opposite spin-polarized signals on the opposite edges of top surface (**Fig. S5**). The orientation of induced spin $S$ seems independent of the positions on a given edge. The upper edge always displays spin-down signals regardless of the position of spin probe Co electrode, while the lower edge always displays spin-up signals. Despite the consistent spin orientation, there exist some difference in the window height of spin voltage loops measured on different Co electrodes of the same edge. The difference may arise from the distinct spin detection efficiency of Co electrodes. Limited by the deposition technique, the spatial inhomogeneity is inevitable for the 3-nm-thick $Al_2O_3$ layer, which could greatly influence the spin detection efficiency of Co electrode. Based on the determined $S$ and current induced momentum $k_e$, we find, in device D, the upper edge inherits left-handed spin-momentum locking, while the lower edge corresponds to right-handed spin-momentum locking.

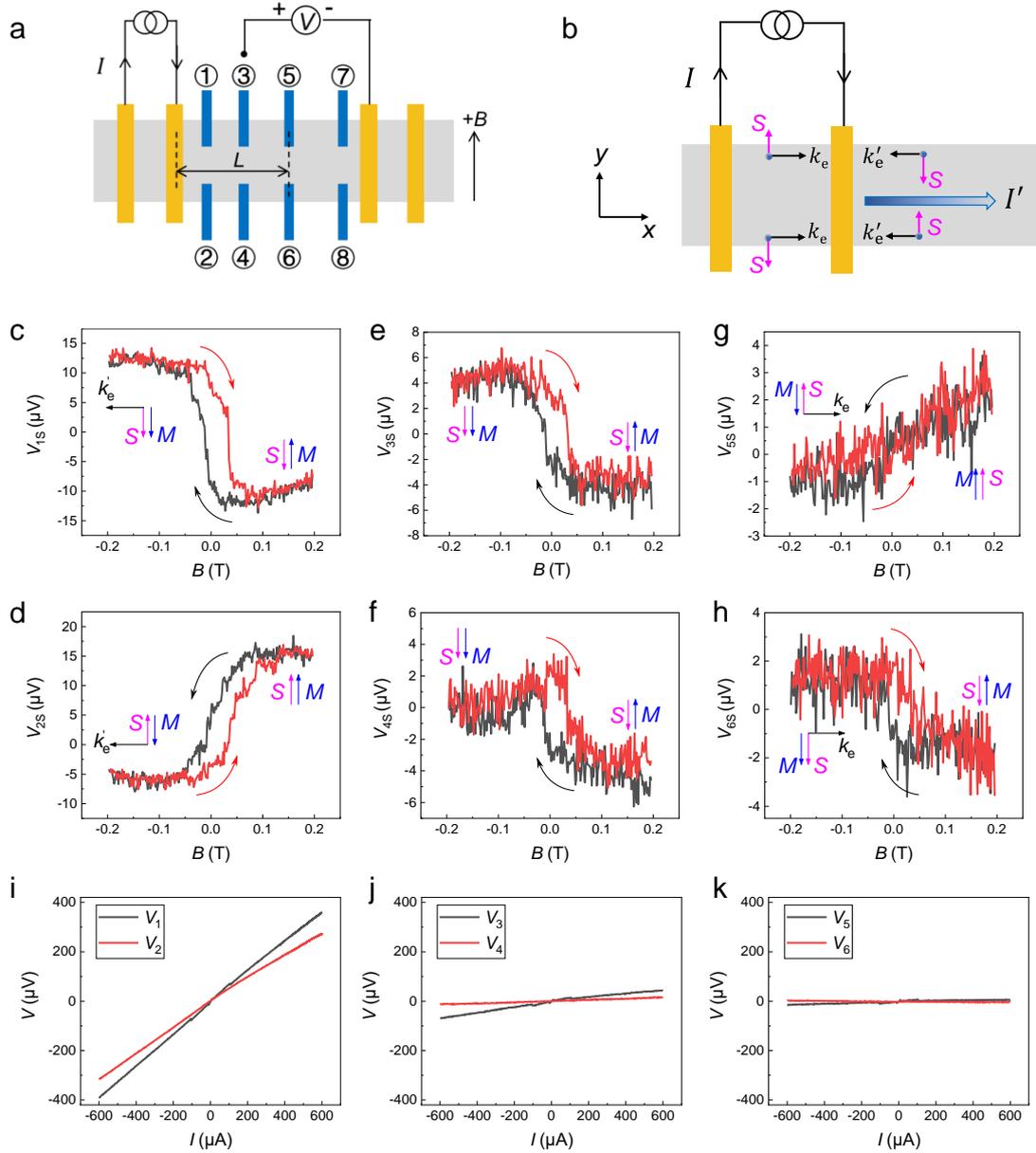

**Figure S6: Nonlocal spin transport measured on the two edges in device D at 1.4 K.**

**a**, Schematic diagram of nonlocal spin detection measurements. The dc bias is injected into the nanoplate via two leftmost Ti/Au electrodes, the voltage is nonlocally measured on each Co electrode with Ti/Au electrode as reference. The voltage measured on Co electrode 1-8 is denoted by $V_1$, $V_2$, …, $V_8$, respectively. The distance $L$ between each pair of Co electrodes and their nearest local Ti/Au electrode is 1.25 μm, 2.25 μm, 3.75 μm and 5.75 μm, respectively.

**b,** Schematic of the induced spin polarization from local current and nonlocal current spreading effect on the two edges. In the local regime, the current $I$ results in a net momentum $k_e$ and thereby a net spin polarization. Due to current spreading effect, the current can nonlocally diffuse away from the local electrode, giving rise to a nonlocal electron momentum $k'_e$ and thereby a net spin polarization.

**c, d,** Spin voltage loops measured on the Co electrode 1 and 2, respectively. The applied local current $I = 600$ μA.

**e, f,** Spin voltage loops measured on the Co electrode 3 and 4, respectively. The applied local current $I = 800$ μA.

**g, h,** Spin voltage loops measured on the Co electrode 5 and 6, respectively. The applied local current $I = 1000$ μA.

**i-k,** Nonlocal current-voltage (*I-V*) curves measured on the Co electrode 1-6 with Ti/Au electrode as reference, respectively.

Nonlocal spin detection measurements have also been performed in device D. Four pairs of Hall-bar-like Co electrode are fabricated as nonlocal spin probes with varied distances from the current electrode (**Fig. S6a**). Contrary to the condition of local spin detection, the nonlocal voltage is detected on the Co electrode away from the source and drain electrodes.

In the nonlocal transport regime, both the spin diffusion process and charge-current-spreading (CCS) effect could produce notable nonlocal spin signals, as shown by **Fig. S6b**. For the spin diffusion process, it comes from the source-drain current induced imbalance between spin-down and spin-up electrons. For the CCS effect, besides local carrier transport between the source and drain electrodes, some carriers would nonlocally spread within mean free path (MFP). The CCS effect is prominent near the current electrode within the MFP of electrons. As shown in **Fig. S6c, d**, when the spin probe Co is close to the local current electrode, spin-down (spin-up) signals are detected on the upper (lower) edge, consistent with the scenario of CCS effect. The linearity of nonlocal *I-V* curves also confirms the dominance of CCS effect (**Fig. S6i**). When the distance $L$ is gradually increased beyond the MFP scale (~1 μm), the CCS effect is greatly quenched, rendering the quick decrease of nonlocal voltage signals (**Fig. S6j, k**). Instead, spin diffusion process gradually comes into prominence, where the reversal of spin voltage loop is observed for each edge (**Fig. S6g, h**).

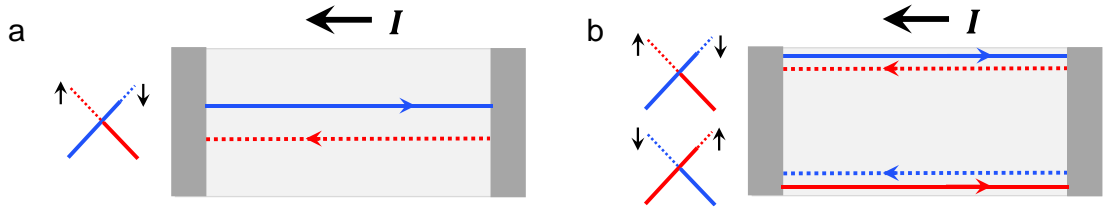

**Figure S7: Comparison of surface state and hinge state spin transport in Cd$_3$As$_2$.**
**a**, Current-induced spin polarizations for surface states. The diagram to the left depicts the population of surface states, where the blue (red) branch corresponds to spin-down (up) electrons. The induced surface spin polarization signal should be spatially independent and almost the same through the same surface.
**b,** Current-induced spin polarizations for hinge states. The diagram to the left depicts the population of hinge states. The current induced spin-polarization directions are opposite for the two neighboring hinges.

**Discussions of possible mechanisms for the opposite spin polarizations**

Several other mechanisms for the observed opposite spin-polarized signals on the two sides of a single nanoplate can be ruled out. First, the topological or Rashba surface states cannot be the origination, because the current-induced surface spin polarization signal should be spatially independent (**Fig. S7**) and almost the same through the same surface [1,2]. Second, the contributions from opposite side wall surfaces are also not consistent with the observations, because the current induced spin orientation of side surface states should be out-of-plane perpendicular to the substrate, while the experiments show that the spin-polarization has prominent components in the substrate plane. The opposite spin-polarized signals are reminiscent of helical edge states that are typical of quantum spin Hall insulators. The presence of quantum spin Hall state is only expected in an ultrathin Dirac semimetal film [3-5], but the thickness of our measured samples is about 100 nm. For the higher-order topological semimetal $Cd_3As_2$ film, current can induce opposite spin polarizations on the two hinges of the same surface. Experimentally, since the Co electrodes are deposited on the top of the nanoplate, the two top hinges would dominantly contribute to the spin signals, while the contribution of bottom hinges can be almost neglected. Thus, opposite spin polarizations are detected on the two edges of nanoplate.

In our work, the 3-nm-thick $Al_2O_3$ layer, inserted between the $Cd_3As_2$ surface and the Co electrode, plays a significant role in manifesting the hinge state spin transport. The oxide layer not only enhances the spin detection efficiency, but also protects the surface and hinge states from the poisoning of ferromagnetic Co electrode. Despite this, the 2D surface states are still inevitably suffering from the degradation of spin polarization due to the finite-angle scattering process [6,7], which could not happen in the 1D hinge channels. Furthermore, the surface Fermi arc states are resulted from fragile topology [8,9], while the 1D hinge states are viewed as topological consequence of bulk Dirac points [8], generally featured with a much higher spin polarization ratio. Therefore, the contrary spin polarizations on opposite edges of nanoplate are observable as a consequence of the dominant contribution of spin-polarized hinge state transport.

# References

[1] Wang A-Q, Xiang P-Z, Ye X-G, et al. Room-temperature manipulation of spin texture in a Dirac semimetal. Phys Rev Applied 2020;14:054044.

[2] Wang A-Q, Xiang P-Z, Ye X-G, et al. Surface engineering of antisymmetric linear magnetoresistance and spin-polarized surface state transport in Dirac semimetals. Nano Lett 2021;21:2026-2032.

[3] Wang Z, Weng H, Wu Q, et al. Three-dimensional Dirac semimetal and quantum transport in $Cd_3As_2$. Phys Rev B 2013;88:125427.

[4] Wang Z, Sun Y, Chen X-Q, et al. Dirac semimetal and topological phase transitions in $A_3Bi$ (A=Na, K, Rb). Phys Rev B 2012;85:195320.

[5] Collins JL, Tadich A, Wu W, et al. Electric-field-tuned topological phase transition in ultrathin $Na_3Bi$. Nature 2018;564:390-394.

[6] Li C-Z, Wang A-Q, Li C, et al. Reducing electronic transport dimension to topological hinge states by increasing geometry size of Dirac semimetal Josephson junctions. Phys Rev Lett 2020;124:156601.

[7] Hakioğlu T. Effect of the electron-phonon interaction on the spin texture in $Bi_{2-y}Sb_ySe_{3-x}Te_x$. Phys Rev B 2019;100:165407.

[8] Wieder BJ, Wang Z, Cano J, et al. Strong and fragile topological Dirac semimetals with higher-order Fermi arcs. Nat Commun 2020;11:627.

[9] Kargarian M, Randeria M, Lu Y-M. Are the surface Fermi arcs in Dirac semimetals topologically protected? Proc Natl Acad Sci U S A 2016;113:8648-8652.